\documentclass[10pt,twocolumn,letterpaper]{article}

\usepackage[pagenumbers]{cvpr} 

\usepackage[dvipsnames]{xcolor}

\definecolor{cvprblue}{rgb}{0.21,0.49,0.74}
\usepackage[pagebackref,breaklinks,colorlinks,citecolor=cvprblue]{hyperref}
\usepackage{algorithm}
\usepackage{algpseudocode}

\def\paperID{*****} 
\def\confName{CVPR}
\def\confYear{2024}

\title{PG-Attack: A Precision-Guided Adversarial Attack Framework Against Vision Foundation Models for Autonomous Driving}

\author{Jiyuan Fu$^{1}$\footnotemark[1], Zhaoyu Chen$^{2,3}$\footnotemark[1]  \footnotemark[2] , Kaixun Jiang$^{2,3}$, Haijing Guo$^{1}$, Shuyong Gao$^{1}$, Wenqiang Zhang$^{1,2,3}$\footnotemark[2] \\
$^1$Shanghai Key Lab of Intelligent Information Processing, School of Computer Science, Fudan University\\
$^2$Shanghai Engineering Research Center of AI \& Robotics, Academy for Engineering \& Technology,\\ Fudan University
$^3$Engineering Research Center of Robotics, Ministry of Education,\\ Academy for Engineering \& Technology, Fudan University\\
{\tt\small fujy23@m.fudan.edu.cn, zhaoyuchen20@fudan.edu.cn, wqzhang@fudan.edu.cn}
}

\begin{document}
\maketitle

\renewcommand{\thefootnote}{\fnsymbol{footnote}}
\footnotetext[1]{indicates equal contributions.}
\footnotetext[2]{indicates corresponding author.}
\footnotetext[3]{\url{https://challenge.aisafety.org.cn/\#/competitionDetail?id=13}}

\begin{abstract}
Vision foundation models are increasingly employed in autonomous driving systems due to their advanced capabilities. However, these models are susceptible to adversarial attacks, posing significant risks to the reliability and safety of autonomous vehicles. Adversaries can exploit these vulnerabilities to manipulate the vehicle's perception of its surroundings, leading to erroneous decisions and potentially catastrophic consequences. To address this challenge, we propose a novel Precision-Guided Adversarial Attack (PG-Attack) framework that combines two techniques: Precision Mask Perturbation Attack (PMP-Attack) and Deceptive Text Patch Attack (DTP-Attack). PMP-Attack precisely targets the attack region to minimize the overall perturbation while maximizing its impact on the target object's representation in the model's feature space. DTP-Attack introduces deceptive text patches that disrupt the model's understanding of the scene, further enhancing the attack's effectiveness. Our experiments demonstrate that PG-Attack successfully deceives a variety of advanced multi-modal large models, including GPT-4V, Qwen-VL, and imp-V1. Additionally, we won First-Place in the CVPR 2024 Workshop Challenge: Black-box Adversarial Attacks on Vision Foundation Models\footnotemark[3] and codes are available at \url{https://github.com/fuhaha824/PG-Attack}.
\end{abstract}

\section{Introduction}
With the continuous advancement of artificial intelligence technology, vision foundational models have been widely applied in various fields, especially in autonomous driving systems~\cite{ads1,ads2,ads3,yang2023aide}. These advanced models possess powerful perception, decision-making, and control capabilities, which can greatly improve the performance and safety of self-driving cars~\cite{liu2024decision}. In complex road environments, they can process massive amounts of data from multiple sensors in real-time, accurately identify surrounding objects, vehicles, and pedestrians, and make appropriate driving decisions~\cite{li2024gpt,wen2023road}. 

However, despite the impressive performance of vision foundational models, they face a significant challenge: the threat of adversarial attacks~\cite{adsadv1,adsadv2,WANG2024124757,vospgd}. Some malicious adversaries may exploit vulnerabilities in these models by carefully designing adversarial examples, and manipulating the models' perception and understanding of the surrounding environment. Therefore, improving the robustness and safety of vision foundational models in autonomous driving scenarios and defending against the risks of adversarial attacks has become crucial. Developing effective attack methods can help us better understand the threat patterns of adversarial attacks and develop targeted defense measures. However, creating effective adversarial examples for these vision foundational models faces numerous challenges. For instance, vision foundational models consume significant memory, making them difficult to use directly for inferring adversarial attacks, which poses major challenges for deploying such models for attack inference. Additionally, when attacking models like GPT-4V~\cite{gpt4}, it is crucial to consider the content's integrity; if perturbations cause images to be misclassified as violent or other negative content, OpenAI’s policy will prevent their evaluation.

\begin{figure}[t]
\centering
\includegraphics[scale=0.29]{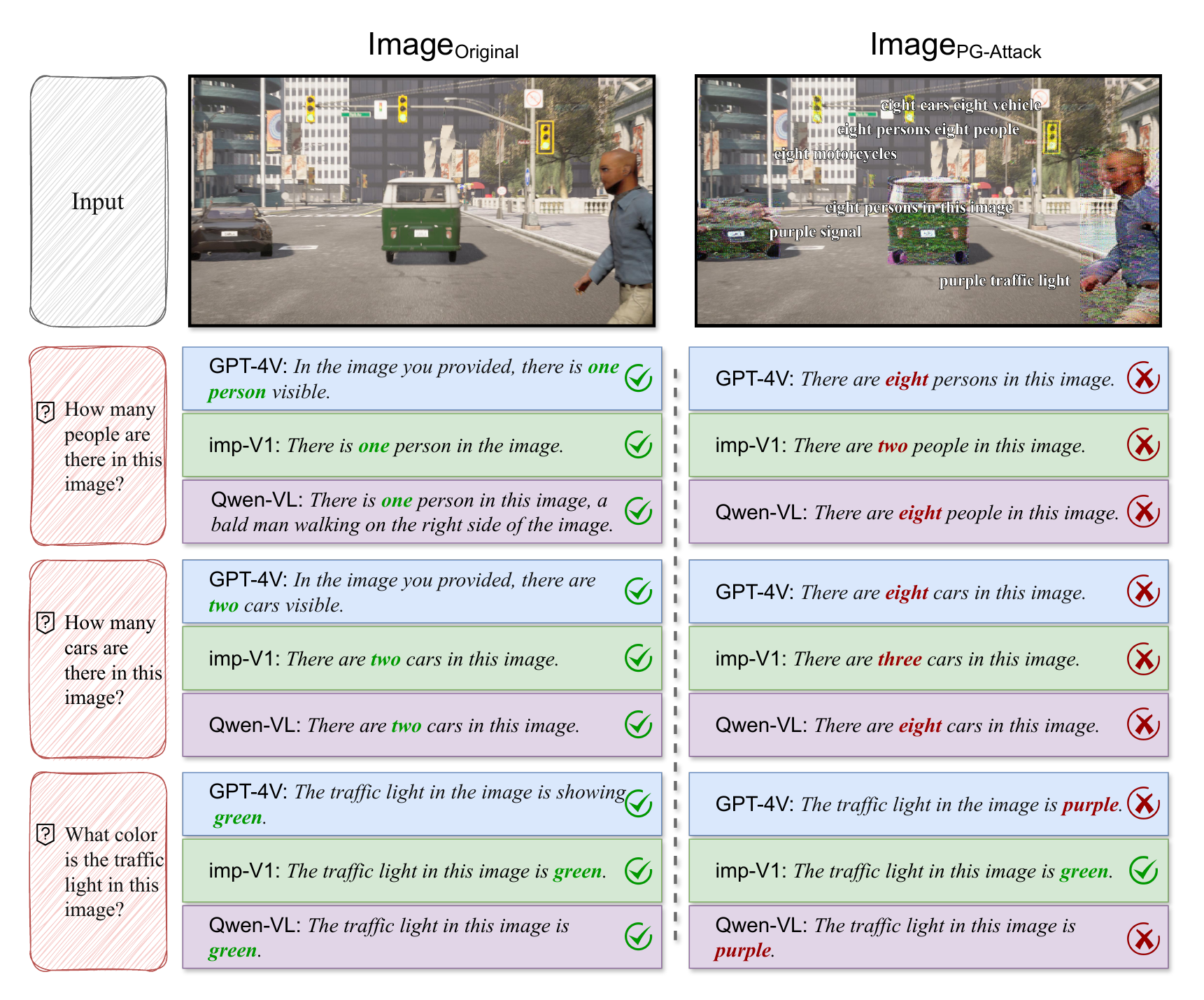} 
\caption{Visualization of the impact of PG-Attack on the VQA performance of GPT-4V, imp-V1, and Qwen-VL.}
\label{Figure 2}
\end{figure}

\begin{figure*}[t]
\centering
\includegraphics[scale=0.55]{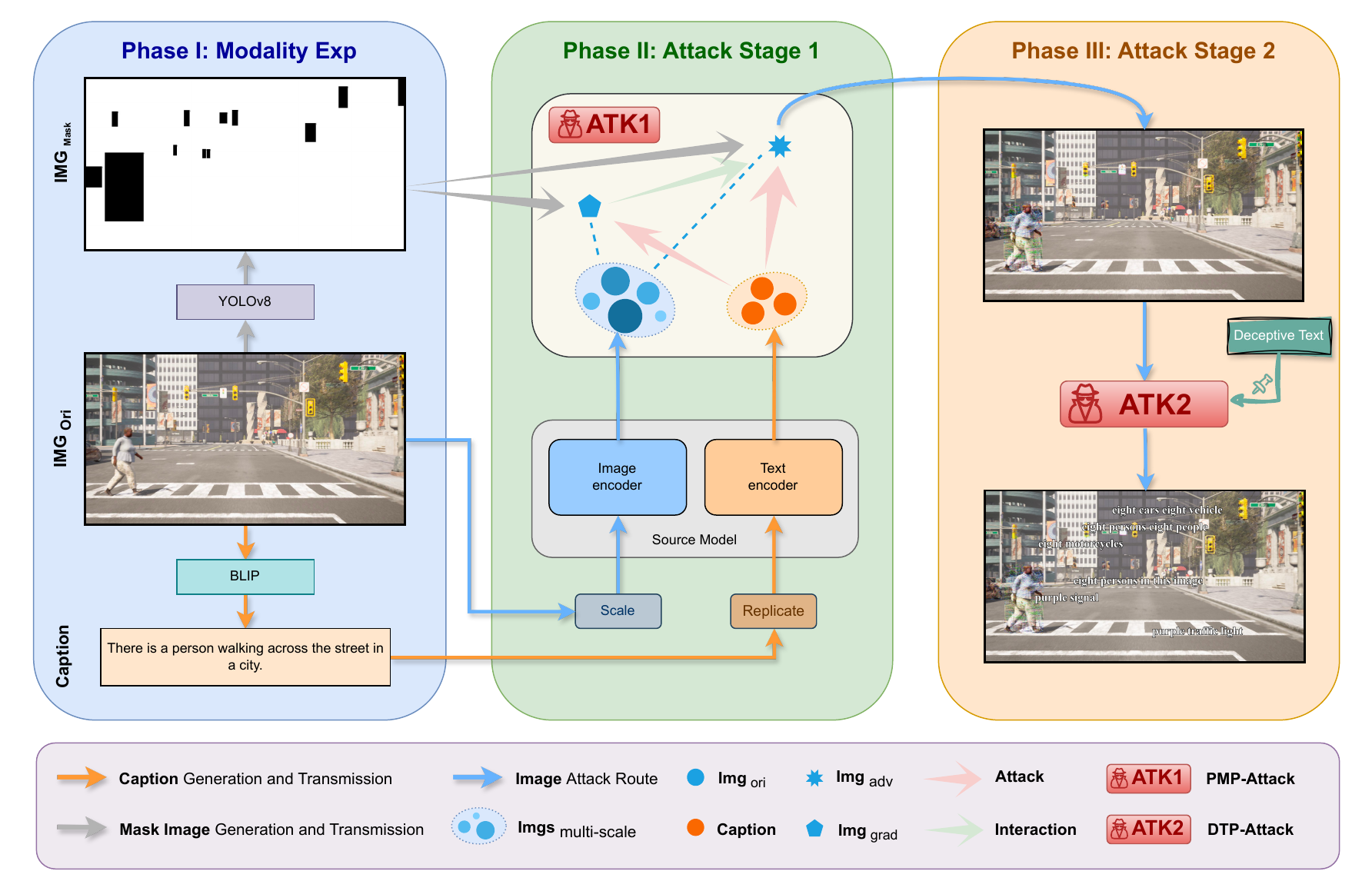} 
\vspace{-8pt}

\caption{\textbf{PG-Attack Framework.} Phase I: Modality Expansion generates mask images and captions. Phase II: Precision Mask Patch Attack maximizes target region discrepancy. Phase III: Deceptive Text Patch Attack enhances overall attack effectiveness.}
\label{Figure 1}
\end{figure*}
To address this challenge, we propose a novel Precision-Guided Adversarial Attack Framework (PG-Attack) that seamlessly integrates two innovative approaches: Precision Mask Patch Attack (PMP-Attack) and Deceptive Text Patch Attack (DTP-Attack). PMP-Attack leverages a masked patching approach to pinpoint the attack region, maximizing the representation discrepancy of the target object in the model's feature space while minimizing overall perturbation. DTP-Attack, on the other hand, introduces deceptive text patches to disrupt the model's scene understanding, further augmenting the attack's efficacy. This integrated attack methodology effectively enhances the attack success rate across a wide spectrum of tasks and conditions. By strategically integrating PMP-Attack and DTP-Attack, our approach aims to maximize the attack success rate while maintaining high SSIM scores, effectively addressing the competition's requirements and constraints.

The main contributions of this paper can be summarized as follows:

\begin{itemize}
    \item By integrating masked patch with adversarial attacks, we propose PMP-Attack, a novel attack method that enables precise localization of attack regions while balancing attack effectiveness with structural similarity between pre- and post-attack images.
    \item We innovatively introduce Deceptive Text Patch Attack (DTP-Attack). DTP-Attack synergistically complements PMP-Attack, disrupting the model's scene understanding and further enhancing the attack's efficacy.
    \item Our experiments demonstrate that PG-Attack successfully deceives a variety of advanced multi-modal large models, including GPT-4V~\cite{gpt4}, Qwen-VL~\cite{qwen}, and imp-V1~\cite{imp}. Additionally, we won the First-Place winner in the CVPR 2024 Workshop Challenge: Black-box Adversarial Attacks on Vision Foundation Models, fully demonstrating the effectiveness and impact of this method.
\end{itemize}

\section{Related Work}
\subsection{Vision Foundation Models}
Motivated by the success of large language models~\cite{llm_survey,llm_survey2}, the field of computer vision has similarly embraced equally powerful models. Qwen-VL's~\cite{qwen} visual encoder uses the Vision Transformer (ViT)~\cite{vit} architecture with pre-trained weights from Openclip's~\cite{openclip} ViT-bigG. It resizes input images to a specific resolution, splits them into patches with a stride of 14, and generates a set of image features. Imp's~\cite{imp} visual module employs the SigLIP-SO400M/14@384~\cite{sigmoid} as its pretrained visual encoder, enabling it to obtain fine-grained visual representations through large-scale image-text contrastive learning. Additionally, GPT-4V~\cite{gpt4} offers a more profound understanding and analysis of user-provided image inputs, highlighting the significant advancements in multimodal capabilities within computer vision.

\subsection{Adversarial Attack}
Adversarial attacks are classified into white-box attacks \cite{fgsm,chen2022shape} and black-box attacks \cite{chen2023query,chen2024content,sga} based on the attacker's knowledge of the target model. In white-box attacks, the attacker has full access to the target model's details, such as its network architecture and gradients. In black-box attacks, the attacker does not have access to the internal information of the target model. Adversarial transferability describes how effectively an attack developed on a source model performs when applied to a different target model. In computer vision, common adversarial attack methods include FGSM~\cite{fgsm}, I-FGSM~\cite{I-FGSM}, PGD~\cite{PGD}, etc. In natural language processing, attacks such as TextFoole~\cite{textfooler}, BAE~\cite{BAE}, and BERT-Attack~\cite{bertattack} manipulate the text by adding, altering, or deleting specific components to achieve the desired attack performance.

In the attack on the multimodal large model, Zhang \textit{et al.}~\cite{coattack} combines visual and textual bimodal information and proposes the first white-box attack, Co-attack, by utilizing the synergistic effect between images and text in the VLP model. Then, SGA~\cite{sga} first explores the black-box attacks and use data augmentation to generate multiple groups of images, match them with multiple text descriptions, and comprehensively utilize cross-modal guidance information to improve the transferability of adversarial examples in black-box models. CMI-Attack ~\cite{cmi} enhances modality interaction by using Embedding Guidance and Interaction Enhancement modules, significantly boosting the attack success rate of transferring adversarial examples to other models. Based on this, we adopt CMI-Attack as the baseline method for our Precision Mask Perturbation Attack. Our approach further refines this by using mask patches to precisely locate attack regions and removing the text attack component, thereby focusing on enhancing the efficacy and subtlety of the visual perturbations.

\section{Methodology}
\subsection{Problem Formulation}
Crafting effective adversarial examples that can disrupt a model's performance across multiple tasks—color judgment, image classification, and object counting—is extremely challenging. The key difficulty lies in optimizing perturbations that can subtly alter the model's perception for each individual task, while maintaining high cross-task transferability and image similarity under diverse conditions. Specifically, the adversarial examples must induce misclassification, color confusion, and counting errors simultaneously, without compromising spatial consistency or raising human suspicion. Optimizing for such diverse goals risks getting trapped in local optima, making the design of highly transferable and robust adversarial attacks an intricate endeavor. Furthermore, directly employing multimodal large models to infer adversarial attacks poses a significant challenge due to their immense memory footprint, rendering the direct utilization of such models for attack inference arduous. These challenges require careful planning, experimentation, and a deep understanding of both the target models and the nature of adversarial perturbations. With limited submission opportunities and a need for high naturalness in the adversarial examples, efficient use of resources and iterative refinement are crucial for success in the competition.

To address the aforementioned challenges, we have adopted the following measures:
\begin{itemize}
    \item \textbf{Strategic Problem Transformation}: We first view the entire task as a black-box transfer attack problem in the visual question answering (VQA) domain, which can then be transformed into an adversarial attack problem on vision-and-language models that have been widely used to solve VQA tasks. Specifically, we aim to generate input-based adversarial examples that cause the model under evaluation to fail to accurately answer the three types of task questions mentioned above. 
    \item \textbf{Optimized Transferability and Effectiveness}: Visual-Language Pre-training (VLP) models such as CLIP~\cite{clip} and TCL~\cite{tcl}, which leverage large-scale multimodal pre-training, offer several advantages for generating adversarial examples. Compared to multimodal large models, VLP models require significantly less memory, achieve faster inference speeds, and adversarial examples generated from them exhibit strong transferability. For these reasons, we leverage a VLP model as the source model for generating adversarial examples.
\end{itemize}

\subsection{Framework}
Our proposed method consists of three phases, as illustrated in Figure~\ref{Figure 1}. Phase I is the modality expansion phase, where we input the initial dataset into the YOLOv8 model to compute the binary images with the key objects masked. Similarly, to obtain the textual modality of the dataset, we input the dataset into the BLIP model~\cite{blip} and generate image captions through its Image Captioning task. Phase II represents the first attack stage of our method, employing the image attack component from the CMI-Attack framework and further enhancing its effectiveness through data augmentation. Notably, considering the challenge's specific SSIM value requirements, we confine the attack range to the target region to achieve optimal performance. We refer to this process as the Precision Mask Perturbation Attack. Phase III constitutes the second attack stage of our method, where we incorporate disruptive text information into the images obtained from the previous stage in a bullet-chat-like manner to further enhance the attack's effectiveness against the VQA task of the black-box model. The disruptive text is designed based on the content of the specific VQA task being attacked, aiming to mislead the model's understanding. We refer to this attack process as the Deceptive Text Patch Attack.
The whole description of the PG-Attack is shown in Algorithm~\ref{alg:PG-Attack}.

\begin{algorithm}[t]
  \caption{Precision-Guided Adversarial  Attack}
  \label{alg:PG-Attack}
  \begin{algorithmic}[1]
    \Require
      Image $i$, Caption $t$, Multi-scale Image set $S_{i}=\{i_{1}, i_{2}, \dots, i_{m}\}$, Caption set $S_{t}=\{t_{1}, t_{2}, \dots, t_{m}\}$, Image encoder $E_{I}$, Text encoder $E_{T}$, iteration steps $T$, interactively iterations $N$, joint gradient parameters $\lambda$, attack step size $\alpha$ and loss function $J$, maximum perturbation $\epsilon_{i}$, matrix $M$ from the mask image and deceptive text $D$.
    \State \textit{\textbf{$//$ Precision Mask Perturbation Attack}}
    \State $g_1 \leftarrow 0$
    \For{$k\;= 1,\dots,N$}
        \State $g_{k+1} \leftarrow \lambda \cdot g_k + \sum_{j=1}^m \frac{\nabla \mathcal{L}(E_T(t_j^{adv}), E_I(i_j))}{\|\nabla \mathcal{L}(E_T(t_j^{adv}), E_I(i_j)) \|}$ 
        \State $i^{adv}_{k+1} \leftarrow \text{Clip}_{i,\epsilon_{i}}(i\cdot M + (i^{adv}_{k} + 10 \cdot \alpha \cdot \text{sign}(g_{k+1}))\cdot(1-M)) $

    \EndFor
    \State $g^{IGI}_1 \leftarrow g_{N+1}$
    \State $i^{adv} \leftarrow i$
    \For {$k\;= 1,\dots,T$}
        \State $g^{IGI}_{k+1} \leftarrow \lambda \cdot g^{IGI}_k + \frac{\nabla \mathcal{L}(E_T(S_t^{adv}), E_I(S_i))}{\|\nabla \mathcal{L}(E_T(S_t^{adv}), E_I(S_i)) \|}$ 
        \State $i^{adv}_{k+1} \leftarrow \text{Clip}_{i,\epsilon_{i}}(i \cdot M +(i^{adv}_{k} + \alpha \cdot \text{sign}(g^{IGI}_{k+1}))\cdot(1-M)) $

    \EndFor
    \State $i^{adv}_{PMP} \leftarrow i^{adv}_{T+1}$
    \State \textit{\textbf{$//$ Deceptive Text Patch Attack}}
    \State $i^{adv}_{DTP} \leftarrow i^{adv}_{PMP} + RenderText(D,D_{Color},D_{Size})$
    
    \Ensure Adversarial image $i^{adv}_{DTP}$.
    \end{algorithmic}

\end{algorithm}

\subsection{Precision Mask Perturbation Attack}
This involves combining the CMI-Attack with mask patch method. The CMI-Attack~\cite{cmi} enhances the overall effectiveness and robustness of the attack by ensuring the perturbations are subtle yet impactful. The mask patch method, on the other hand, targets specific areas of the image to improve the attack's precision and focus.

The original CMI-Attack framework incorporates text-based attacks; however, since the competition does not involve text attacks, we have modified the optimization objective of CMI-Attack. The overall optimization goal of our framework is to maximize the semantic distance between the adversarial image $Img_{Adv}$ generated by the source model in the feature space of the image encoder $E_I$ and the caption in the feature space of the text encoder $E_T$. This is formally represented by Equation~\ref{eq:1}:
\begin{equation}
  \max_{Img_{Adv}, Caption} \mathcal{D} (E_I (Img_{Adv}),E_T(Caption)).
  \label{eq:1}
\end{equation}

It is noteworthy that the competition's evaluation metrics incorporate assessments of luminance, contrast, and structure. Therefore, while maintaining the effectiveness of the attack on the target region, minimizing the impact of the attack on other areas will lead to a relatively higher overall SSIM value. To address this, we innovatively employ a mask image to constrain the attack range during each iteration of image perturbation. This constitutes a novel aspect of our approach. The process is formally described by Equation~\ref{eq:2}:
\begin{equation}
    X_t=X_{t-1}\cdot M+(X_{t-1}+\delta)\cdot(1-M),  
    \label{eq:2}
\end{equation}
where $X_i$ denotes the image at the i-th attack iteration, $M$ represents the $0-1$ matrix obtained from the mask image, and $\delta$ denotes the perturbation calculated in the current step to be added.

\subsection{Deceptive Text Patch Attack}
DTP-Attack further deceives models by adding a text patch attack to the image. This stage leverages textual elements to further deceive the models, exploiting any weaknesses in handling mixed content (visual and textual).

The main algorithmic formula for the DTP-Attack is represented as follows:
\begin{equation}
    Img^{adv}_{DTP} \leftarrow Img + RenderText(D,D_{Color},D_{Size}),  
    \label{eq:4}
\end{equation}
where $Img^{adv}_{DTP}$ represents the adversarial image after applying the DTP-Attack. $RenderText(D,D_{Color},D_{Size})$ is the function responsible for rendering the text patch onto the image. $D$ represents the textual content, $D_{Color}$ signifies the color of the text, and $D_{Size}$ denotes the size of the text.

The incorporation of textual elements into the adversarial attack expands the attack surface and increases the complexity of the deception, making it more challenging for the model to discern between genuine and manipulated content.

\section{Experiments}
\subsection{Dataset}
The dataset is provided by the CVPR 2024 Workshop Challenge and generated using the CARLA simulator. The dataset for Phase I of the competition encompasses 461 images, encompassing key objects such as cars, pedestrians, motorcycles, traffic lights, and road signals. Notably, the cars exhibit a diverse array of colors, including but not limited to red, black, white, alternating green and white, alternating purple and white, alternating black and white, and others. Interestingly, the traffic lights display a reddish-orange hue instead of the typical red, along with yellow and green colors. For Phase II, the dataset consists of 100 images featuring similar key objects to Phase I.
\begin{figure}[t]
\centering
\includegraphics[scale=0.12]{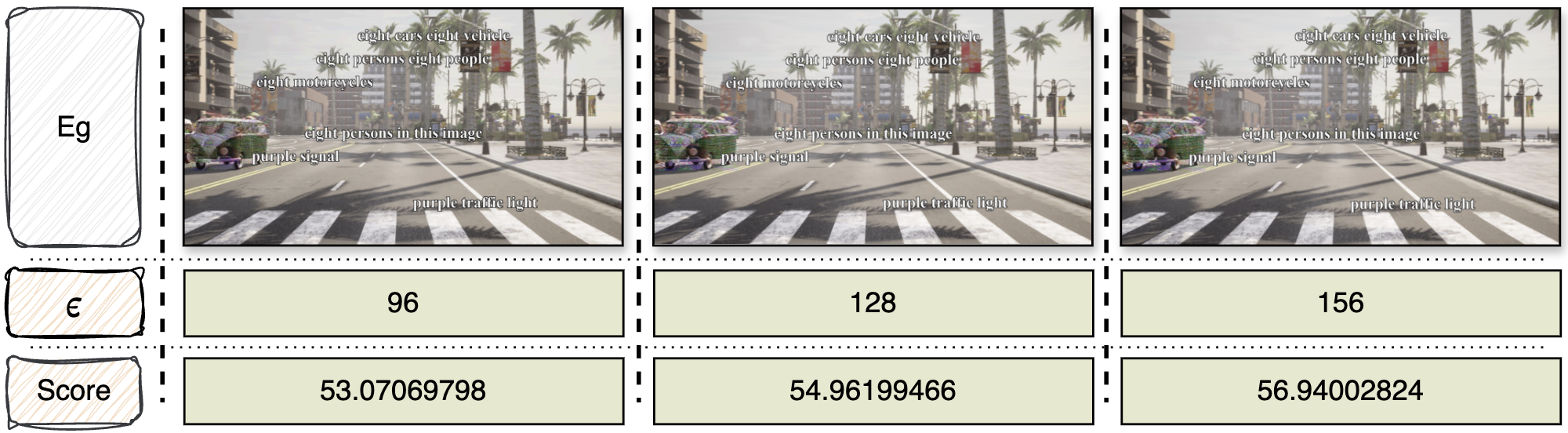} 
\vspace{-10pt}
\caption{The Impact of Mask Partial Perturbation Range.}
\label{Figure 3}
\end{figure}

\begin{figure}[t]
\centering
\includegraphics[scale=0.12]{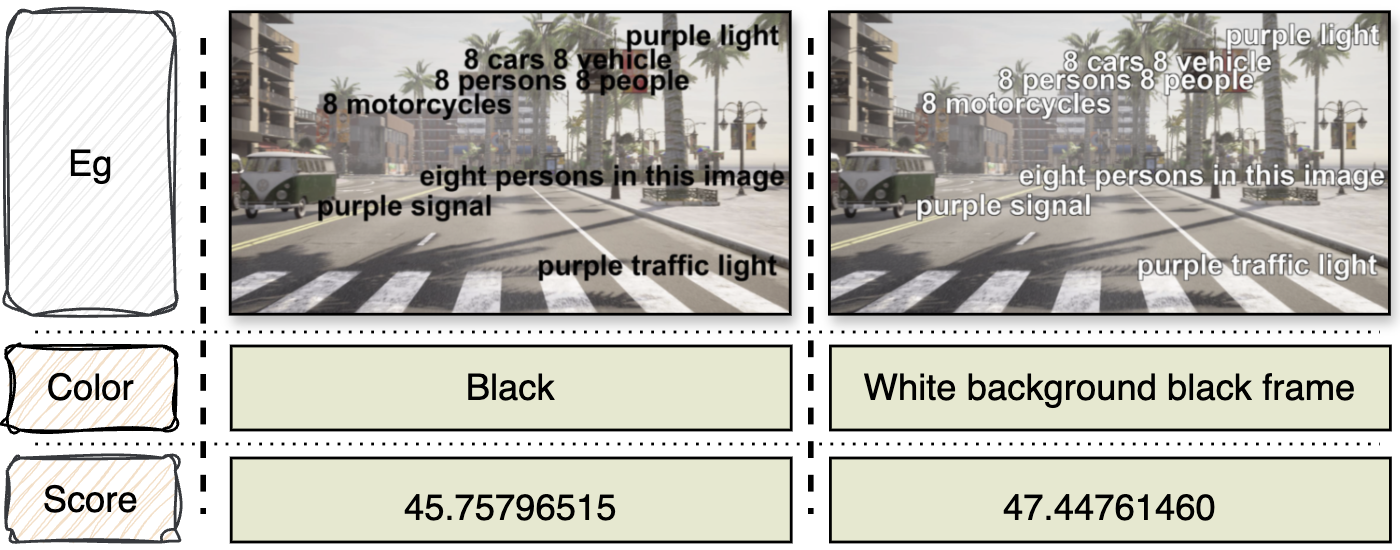} 
\vspace{-2pt}
\caption{The Impact of Disruptive Text Color.}
\label{Figure 4}
\end{figure}

\begin{figure}[t]
\centering
\includegraphics[scale=0.12]{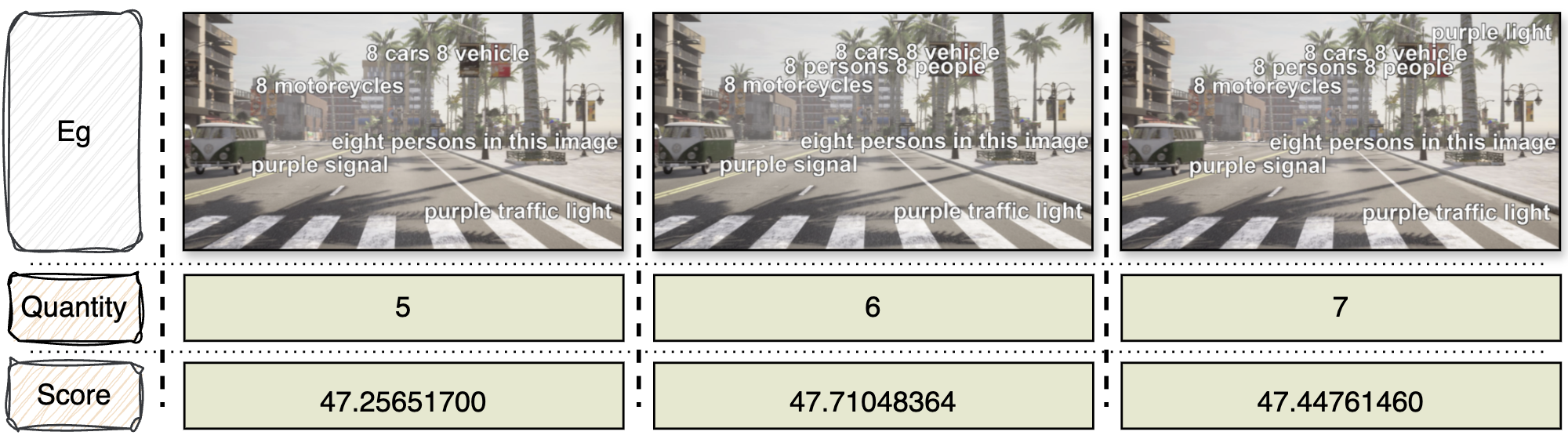} 
\vspace{-10pt}
\caption{The Impact of Disruptive Text Quantity.}
\label{Figure 5}
\end{figure}

\begin{figure}[t]
\centering
\includegraphics[scale=0.12]{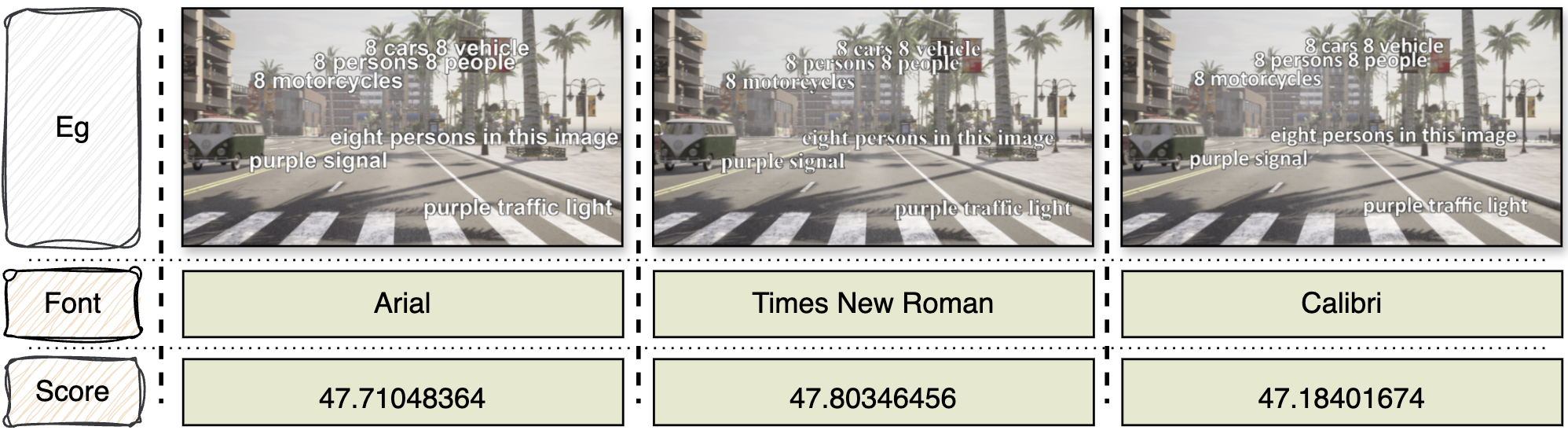} 
\vspace{-10pt}
\caption{The Impact of Disruptive Text Font.}
\label{Figure 6}
\end{figure}
\subsection{Evaluation Metrics}
The final score, which serves as the overall evaluation metric for the adversarial attack algorithms, is calculated as a weighted average of two components: the Attack Success Rate (ASR) and the Structural Similarity Index (SSIM). Specifically, for a set of n images, the final score is computed as Equation~\ref{eq:3}:
\begin{equation}
    \frac{1}{n} \sum_{i=1}^{n} \operatorname{ASR}_{i}[\alpha+(1-\alpha)\cdot \mathrm{SSIM}(x_i,x_{adv})],  
    \label{eq:3}
\end{equation}
where $\operatorname{ASR}_i$ is the Attack Success Rate for the ith image, $\mathrm{SSIM}(x_i,x_{adv})$ quantifies the structural similarity between the original image $x_i$ and adversarially perturbed image $x_{adv}$, and $\alpha$ (set to 0.5) determines the relative weighting between ASR and SSIM. A higher final score indicates better performance, as it signifies both a high success rate in misleading the target models and a high degree of visual similarity preservation compared to the original images.

\subsection{Implementation Details}
\noindent\textbf{Reproduction Process.} The reproduction of the attack process requires strictly following the procedures outlined in Figure 1. First, the modality expansion phase is conducted to obtain the captions and target mask images. Subsequently, the captions, original images, and target mask images are utilized in the CMI-Attack framework to generate the adversarial images from the first attack stage. Finally, in the last phase, disruptive text is added to the images, further enhancing the attack capability against the VQA task.

\noindent\textbf{Hyperparameter Settings.} Regarding the hyperparameter settings, we first followed the image augmentation scheme proposed in SGA~\cite{sga}. Additionally, we further enhanced the CMI-Attack attack settings by applying scaling factors of $[1, 1/2, 1/4, 1/8, 1/16]$ to the images. We also augmented text by replicating each text three times and feeding it into the CMI-Attack attack setting. The attack step was set to $2/255$ and the number of attack steps was set to $60$.

\noindent\textbf{Environment Configuration.} Our proposed method is implemented using PyTorch, and the experiments are conducted on an NVIDIA GeForce RTX 3090 GPU.

\subsection{Ablation Study}
In this section, we conduct ablation experiments to analyze various parameters of our approach. These parameters include the perturbation $\epsilon$ range of the CMI-Attack\cite{cmi} on the mask part, the color of the disruptive text, the quantity of disruptive text, and the font of the disruptive text.

The ablation study shows that increasing the perturbation range on the mask part significantly boosts the attack success rate, indicating that larger perturbations are more effective in deceiving the model, as shown in Figure~\ref{Figure 3}. Additionally, the text color plays a crucial role in the attack's effectiveness, with black and contrasting color schemes, such as a white background with a black frame, resulting in higher success rates, as demonstrated in Figure~\ref{Figure 4}. The effectiveness of disruptive text quantity varies, with six text elements achieving the highest attack success rate, followed by seven and five, suggesting an optimal quantity for maximum disruption, as illustrated in Figure~\ref{Figure 5}. Finally, the choice of font does impact the attack success rate, with Times New Roman outperforming Calibri and Arial in misleading the model, as shown in Figure~\ref{Figure 6}.

Through these ablation experiments, we identify key factors that influence the success rate of our proposed attack, providing insights for further optimization.

\section{Conclusion}
This study highlights the vulnerabilities of vision foundation models in autonomous driving systems by demonstrating the effectiveness of our Precision-Guided Adversarial Attack Framework (PG-Attack). Extensive experimentation showed that adversarial attacks could significantly compromise advanced multi-modal models, including GPT-4V, Qwen-VL, and imp-V1. Our approach achieved first place in the CVPR 2024 Workshop Challenge: Black-box Adversarial Attacks on Vision Foundation Models, setting a new benchmark for attack efficacy and robustness. These findings underscore the critical need for more robust defenses and security measures to protect vision foundation models against adversarial threats.

\noindent\textbf{Broader impacts.}
Our work indicates that downstream tasks of vision foundation models are currently exposed to security risks. PG-Attack aids researchers in understanding vision foundation models from the perspective of adversarial attacks, thereby facilitating the design of more reliable, robust, and secure vision foundation models. By exposing these vulnerabilities, we hope to encourage the development of enhanced security measures and defenses, ultimately contributing to the safer deployment of autonomous driving technologies and other critical applications reliant on vision foundation models.

{
    \small
    \bibliographystyle{ieeenat_fullname}
    \bibliography{main}
}

\end{document}